\documentclass[10pt,sigconf,letterpaper]{acmart} 

\makeatletter
\def\@ACM@checkaffil{
    \if@ACM@instpresent\else
    \ClassWarningNoLine{\@classname}{No institution present for an affiliation}%
    \fi
    \if@ACM@citypresent\else
    \ClassWarningNoLine{\@classname}{No city present for an affiliation}%
    \fi
    \if@ACM@countrypresent\else
        \ClassWarningNoLine{\@classname}{No country present for an affiliation}%
    \fi
}
\makeatother
\usepackage[english]{babel}
\usepackage{algorithmic}
\usepackage{graphicx}
\usepackage{textcomp}
\usepackage{xcolor}
\usepackage{booktabs}
\usepackage{makecell}
\usepackage{tabularx}
\usepackage{multirow}
\usepackage{csquotes}
\usepackage{xspace}
\usepackage{url}
\usepackage{numprint}

\usepackage{stfloats}

\author{Salim Chouaki}
\affiliation{%
  \institution{LIX, CNRS, Inria, Ecole Polytechnique, Institut Polytechnique de Paris}
}
\email{salim.chouaki@inria.fr}

\author{Oana Goga}
\affiliation{%
  \institution{LIX, CNRS, Inria, Ecole Polytechnique, Institut Polytechnique de Paris}
}
\email{oana.goga@cnrs.fr}

\author{Hamed Haddadi}
\affiliation{%
  \institution{Imperial College London, Brave Software}
}
\email{h.haddadi@imperial.ac.uk}

\author{Peter Snyder}
\affiliation{%
  \institution{Brave Software}
}
\email{pes@brave.com}

\copyrightyear{2023} 
\acmYear{2023} 
\setcopyright{acmlicensed}\acmConference[IMC '23]{Proceedings of the 2023 ACM Internet Measurement Conference}{October 24--26, 2023}{Montreal, QC, Canada}
\acmBooktitle{Proceedings of the 2023 ACM Internet Measurement Conference (IMC '23), October 24--26, 2023, Montreal, QC, Canada}
\acmPrice{15.00}
\acmDOI{10.1145/3618257.3624823}
\acmISBN{979-8-4007-0382-9/23/10}

\newenvironment{myquote}%
  {\list{}{\leftmargin=0.15in\rightmargin=0.1in}\item[]}%
  {\endlist}

\usepackage{xspace}

\newcommand{\IE}{i.e.,}
\newcommand{\EG}{e.g.,}

\newcommand{\SP}{StartPage\xspace}
\newcommand{\DDG}{DuckDuckGo\xspace}
\newcommand{\EL}{EasyList\xspace}
\newcommand{\EP}{EasyPrivacy\xspace}

\begin{document}

\title{Understanding the Privacy Risks of Popular Search Engine Advertising Systems}

\ccsdesc[500]{Security and privacy~Privacy protections}
\ccsdesc[500]{Networks~Network measurement}

\ccsdesc[500]{Security and privacy~Privacy protections}
\ccsdesc[500]{Networks~Network measurement}

\keywords{Search engines, advertising systems, cross-site tracking, \\privacy, measurement.} 

\begin{abstract}

We present the first extensive measurement of the privacy properties of the advertising systems used by privacy-focused search engines. We propose an automated methodology to study the impact of clicking on search ads on three popular \textit{private} search engines which have advertising-based business models: StartPage, Qwant, and DuckDuckGo, and we compare them to two dominant data-harvesting ones: Google and Bing. We investigate the possibility of third parties tracking users when clicking on ads by analyzing first-party storage, redirection domain paths, and requests sent before, when, and after the clicks. 

Our results show that privacy-focused search engines fail to protect users' privacy when clicking ads. Users' requests are sent through redirectors on 4\% of ad clicks on Bing, 86\% of ad clicks on Qwant, and 100\% of ad clicks on Google, DuckDuckGo, and StartPage. Even worse, advertising systems collude with advertisers across all search engines by passing unique IDs to advertisers in most ad clicks. These IDs allow redirectors to aggregate users' activity on ads' destination websites in addition to the activity they record when users are redirected through them. Overall, we observe that both privacy-focused and traditional search engines engage in privacy-harming behaviors allowing cross-site tracking, even in privacy-enhanced browsers.

\end{abstract}

\maketitle


\section{Introduction}

Privacy-focused search engines such as \DDG, \SP, and Qwant~\cite{ddg, startpage, qwant} promote a strategy of respecting users' privacy and promise not to track users' search and browsing behavior, all while delivering relevant search results. 
However, private search engines rely on advertising for revenue, and use traditional advertising platforms to deliver ads: \DDG and Qwant use Microsoft's advertising system, while \SP uses Google's advertising system. 
These search engines are often ambiguous on the privacy properties of the ads that appear on their search page, and their consequent privacy properties remain unexplored to the best of our knowledge.

In this work, we aim to fill this gap by conducting the first study of the privacy properties of the advertising systems of three major privacy-focused search engines: \DDG, \SP, and Qwant, and how they compare to more popular search engines: Bing and Google. We investigate the privacy properties of these search engines when they: (i) present search ads to users, (ii) when a user clicks on an ad, and (iii) when the user lands on the advertiser's page.

We implement an automated measurement methodology to measure if and how users can be re-identified (hence, their privacy is compromised) when clicking on search ads on each search engine (see Section~\ref{sec:metho}). 
We build an open-source implementation of this methodology in the form of a Puppeteer-based pipeline that simulates search queries and ad clicks. 
We apply this crawling methodology to the five search engines, providing a full dataset with visited websites, cookies created, locally stored values, and web requests to search engines' servers and/or other third parties when clicking ads. 
We use filter rules from several major open-source lists to detect web requests to online trackers, and we propose a methodology to differentiate user identifiers from non-tracking values in query parameters and cookies values.

We then present in Section~\ref{sec:results} a systematic analysis of our dataset to investigate privacy harms before clicking an ad, during clicking an ad, and after clicking an ad and reaching the advertiser's website. 
We find that users' privacy is not harmed \emph{until} users click on an ad. 
Privacy-focused search engines do not appear to attempt to re-identify users across visits or queries and do not include resources from, or make network requests to known trackers.
However, we find that users' privacy is compromised by \textbf{all} studied search engines in various ways once users click on an ad.

Disappointingly, we find that all search engines record additional information about the user and/or the users' clicks after the user has clicked on an ad. 
Private search engines capture data related to the clicked ad, including the ad provider, destination URL, and the ad's position within the search results page, along with the user's browsing data, such as the search query, device type, and browser language. Private search engines do not store user identifiers upon ad clicks, in contrast to traditional search engines that record user identifying values.
Furthermore, we find that all search engines 
engage in navigation-based tracking. 
Navigation-based tracking refers to tracking techniques that are redirecting users through one or more redirectors when navigating from one website to another in order to share user information across sites~\cite{randall2022measuring}. 
Navigation-based tracking does not require third-party cookies and can be used to circumvent browsers' privacy protections from cross-site tracking using partitioned cookies storage.
Alarmingly, we observe that privacy-focused search engines engage in more navigation-based tracking than non-privacy-focused ones:  
We observe navigational tracking on 4\% ad clicks on Bing, on 100\% ad clicks on Google, on 100\% ad clicks on \DDG, on 86\% ad clicks on Qwant, and on 100\% ad clicks on StartPage.

On the destination page, we check whether the search engine requires advertisers to abide by privacy-respecting practices by measuring whether advertisers include trackers or other known privacy-harming resources. 
We found that 93\% of ads destination pages (across all five search engines) included tracker and privacy-harming resources. 
Finally, we check whether search engines or redirectors aid advertisers in profiling visitors by measuring the data they receive in the form of user-describing query params. 
We find that advertisers receive user identifiers in 68\%, 92\%, and 53\% of cases for \DDG, \SP, and Qwant, respectively. 
This practice, known as UID smuggling, enables redirectors to aggregate more user behavior data if they have scripts on the ads' destination websites and they store the user-identifying parameters they receive. 
Notably, in the case of private search engines, the user-identifying parameters are not set by the search engine but by the redirectors encountered between the search engine's and the advertiser's sites.

Our results indicate that private search engines' privacy protections do not sufficiently cover their advertising systems. Although these search engines refrain from identifying and tracking users and their ad clicks, the presence of ads from Google or Microsoft subjects users to the privacy-invasive practices performed by these two advertising platforms. 
When users click on ads on private search engines, they are often identified and tracked either by Google, Microsoft, or other third parties, through bounce tracking and UID smuggling techniques. 
Particularly, advertisers receive unique user identifiers through query parameters in most ad clicks, which can enable cross-site tracking even in privacy-enhanced browsers that block third-party cookie tracking. 

\section{Background}

This section briefly discusses the policies 
of the main search engines alongside popular tracking approaches.

\subsection{Private search engines}
\label{sec:background:se}
We study the two dominant search engines that rely on user tracking for personalized search results and advertisements, namely Google and Bing, and three of the most popular privacy-branded search engines that provide users with non-personalized results and ads: \DDG, \SP, and Qwant~\cite{nord_vpn_best_private_ses, brave_best_private_ses}. Private search engines can either build their own independent search indexes or use big tech search engines like Bing, Google, or Yahoo to provide search results. Both types of private search engines claim not to store users' search histories and not to collect nor share tracking and personal data. We now describe the advertising systems employed by the different private search engines and present a summary of their data-sharing policies outlined in their respective \emph{About} pages.

\vspace{2mm}
\noindent \textbf{\DDG} is a standalone search engine that maintains and uses its own search index alongside other indexes, such as Bing's, to provide search results~\cite{ddg_privacy_anonymzed_results}. \DDG relies on Microsoft's advertising system but only serves ads based on the search results and not the behavioral profiles of users~\cite{ddg_advertising}:

\begin{myquote}
\small{{\em 
"search ads on \DDG are based on the search results page you're viewing instead of being based on you as a person"}}
\end{myquote}

When clicking an ad on \DDG, the user is redirected to the ad's landing page through Microsoft Advertising's platform. \DDG claims Microsoft does not store ad-click behaviors from \DDG for purposes other than accounting and does not associate ad-clicks with users' profiles~\cite{ddg_microsoft_advertising}:
\begin{myquote}
\small{{\em 
"When you click on a Microsoft-provided ad that appears on \DDG, Microsoft Advertising does not associate your ad-click behavior with a user profile. It also does not store or share that information other than for accounting purposes." }}
\end{myquote}
This implies that Microsoft can, though currently chooses not to, link the ad-click to an existing Microsoft user profile. The privacy policy is signed by both \DDG and Microsoft.

\vspace{2mm}
\noindent \textbf{Qwant} is a standalone EU-based search engine that allows users to access online resources without being tracked nor profiled~\cite{qwant_legal_information}. Qwant relies on Microsoft's advertising system to deliver ads in their search results pages. Although Qwant reports transmitting \emph{some information} concerning search queries to Microsoft to enable the latter to present pertinent advertisements, it remains unclear which specific information is shared. In addition, to detect fraud, Qwant uses a specialized service offered by Microsoft, which has access to the user's IP address and the browser "User-Agent". Qwant assures that this service does not have access to the search query, which is sent to another service that does not know the IP address of the user~\cite{qwant_legal_information}. 

Unlike \DDG, which also uses Microsoft advertising, we did not find any mention to ad-click 
information on Qwant's privacy policy. They do not mention whether Microsoft stores this data and for what purposes they use it.

\vspace{2mm}
\noindent \textbf{\SP} is a meta-search engine that allows users to obtain non-personalized search results from Google's search index while protecting their privacy. \SP relies on Google AdSense to show ads to users. 
According to \SP's privacy policy, the search engine serves strictly non-personalized ads since it does not share any identifiable information with Google. Therefore, ads displayed on the search results page are solely based on the user's search query~\cite{startpage_privacy_policy}.

Regarding ad-click behavior data, the privacy policy does not make any reference to whether Google tracks or profiles users based on this information. Nevertheless, \SP emphasizes that by clicking on an ad, users leave the protection of \SP's privacy policies and become subject to the 
practices of the website they are redirected to~\cite{startpage_can_i_advertise}.

\begin{myquote}
\small{{\em 
"By clicking on an ad, like any other external website you click on after performing a \SP search, you leave the privacy protection of \SP and are subject to those websites' data collection policies." }}
\end{myquote}

\subsection{Cross-site tracking}
Cross-site tracking refers to the practice of following a user across multiple first-party websites and associate their browsing activities to a unique identifier. Web tracking practices require first-party websites (e.g. the content providers) to share data about a user's activity with third parties (the trackers). 
Online tracking has been traditionally implemented through browser cookies. However, due to increasing adoption of cookie-blocking browsers and extensions, and the push on adopting partitioned cookies storage on web browsers, more and more trackers started to rely on navigational tracking techniques. We next discuss 
how these 
techniques work.

\subsubsection{Cookie tracking} 
\begin{figure*}[t!]
  \centering
  \includegraphics[width=0.8\linewidth]{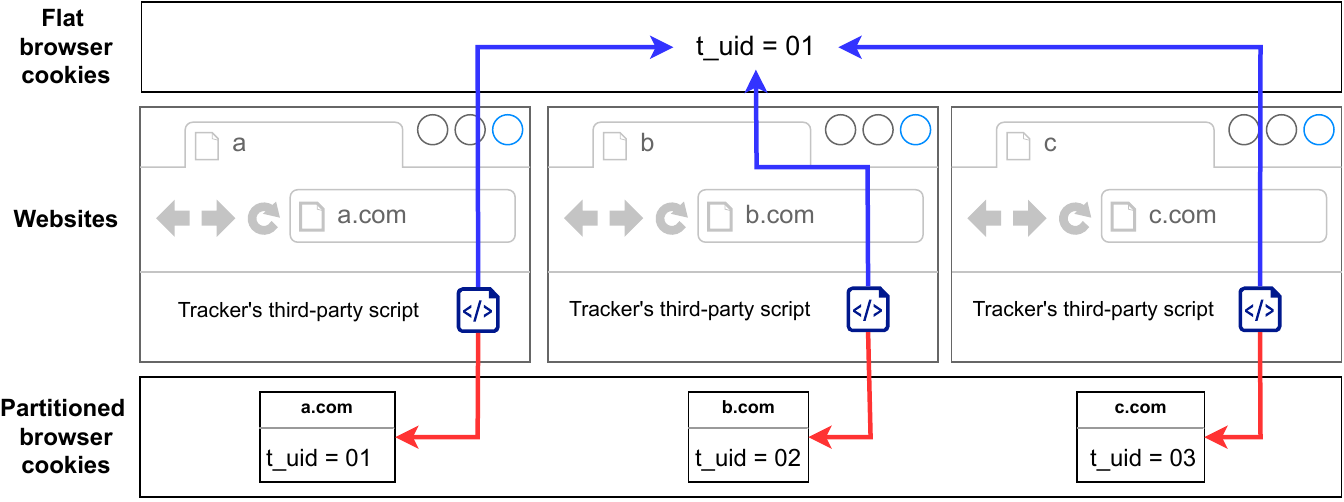}
  \caption{Cookie tracking in flat vs. partitioned cookies storage.}
  \label{fig:cookies_tracking}
\end{figure*}

To enable cross-site cookie tracking, whenever a user visits a first-party website, the website makes a request to the third-party website (the tracker). This allows the tracker to set a cookie, which will identify the user and will be associated with the browsing activity of the user.  
For example, when the user visits a website \emph{A} that makes a request to the tracker \emph{T}, the tracker associates the cookie identifier of the user with the fact that the user visited website \emph{A} (see Figure~\ref{fig:cookies_tracking}). Later, when the user visits website \emph{B}, which also makes a request to the tracker \emph{T}, the tracker will be able to associate the cookie identifier of the user with the fact that the user visited website \emph{B}. Hence, the tracker will be able to know that the user visited websites \emph{A} and \emph{B}. 

This was initially possible because browsers had a common cookie storage containing all cookies, and trackers could read their corresponding cookies regardless of which first-party website allowed the tracker cookie to be set (see Figure~\ref{fig:cookies_tracking}). 
However, several browsers, such as Safari, Firefox, and Brave, have implemented partitioned storage to prevent using cookies for cross-site tracking~\cite{randall2022measuring}. These browsers use a partitioned cookies storage with a hierarchical namespace where a tracker accesses a different storage area on each website that loads it, preventing trackers from matching or assigning the same identifiers to users across multiple websites. Hence, cross-site tracking based on cookies can no longer be performed on these browsers. Chrome -the most used web browser- is in the process of testing partitioned cookies storage but does not use it by default~\cite{google_sandbox, google_sandbox_2}.

\subsubsection{Navigational tracking}
\label{sec:background:nav}

Navigational tracking refers to tracking techniques that use one or more URL navigations to share user information across sites. Navigational tracking does not require third-party cookies and can be used to circumvent browsers' privacy protections from cross-site tracking using partitioned cookies storage.

\begin{figure*}[t!]
  \centering
  \includegraphics[width=0.80\linewidth]{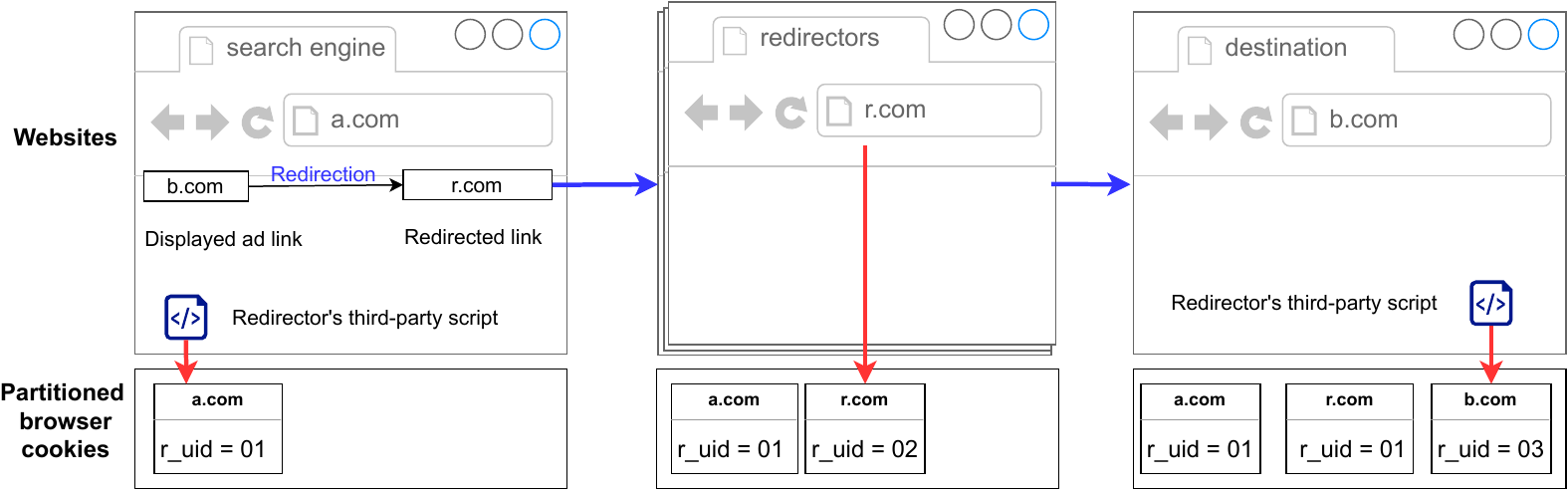}
  \caption{Bounce tracking.}
  \label{fig:bounce_tracking}
\end{figure*}

\vspace{2mm}
\textbf{Bounce tracking} is a navigational tracking technique that refers to redirecting users through one or more redirectors when navigating from one website to another. To allow this, a website \emph{A} containing links to another website \emph{B} does not directly link to the target \emph{B} but instead links to an intermediary \textit{redirector} (R)--the tracker (see Figure~\ref{fig:bounce_tracking}). 
When users click on a link on website \emph{A}, they are taken to the redirector first, which then redirects them to the intended destination (website \emph{B}) or other intermediary redirectors. 
The website \emph{A} can directly change the actual link of the destination (b.com) to a redirection link (r.com), or a redirector's third-party script can do it. On its turn, the redirector can change the destination link again and send it further to other redirectors. Hence, from the link in the ad on the website \emph{A}, one cannot know all the different redirectors the users will pass through when they click on an ad. We call the \emph{redirection path} all the websites a user navigates through to arrive from \emph{A} to \emph{B}.  
Since, from a browser perspective, the redirector is the first-party domain, it can read or set cookies in its own partition~\cite{Koop_indepth_evaluation}. In the following, we describe what data redirectors can infer according to the redirector's behavior. 

\begin{trivlist}
\vspace{-2mm}
\item \hspace{2mm} (1) If the redirector does not set a first-party cookie, it will only know that a user went from website \emph{A} to website \emph{B} and will not be able to link this to other user browsing activities. 

\item \hspace{2mm} (2) If the redirector sets a first-party cookie, it will be able to aggregate all the activity of the user that is redirected through it (either from website A or other websites that use it as a redirector), hence, it will allow cross-site tracking. 

\item \hspace{2mm} (3) If the redirector also sets third-party cookies on websites \emph{A} and \emph{B}, it will not be able to link the activity of the user on website \emph{A} with the activity of the user on website \emph{B}, and with the activity of the user that goes through its own site (through redirects) since they do not share the same user ID~\cite{randall2022measuring}. Hence, while bounce tracking allows to a certain degree, cross-site tracking, it does not have the same coverage as the traditional 
third-party cookie tracking. 
\end{trivlist}

\begin{figure*}[t!]
  \centering
  \includegraphics[width=0.8\linewidth]{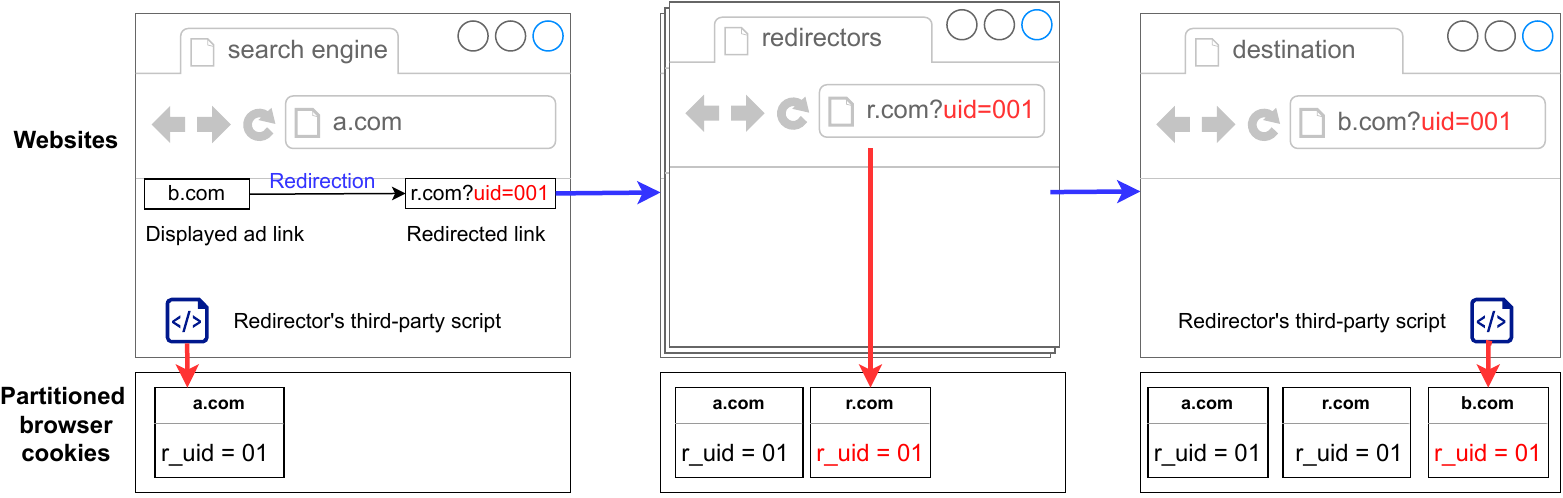}
  \caption{UID smuggling}
  \label{fig:uid_smuggling}
  
\end{figure*}

\textbf{UID smuggling} is a navigational tracking technique that modifies users' navigation requests by adding information to the navigation URLs in the form of query parameters. In addition, similar to bounce tracking, UID smuggling may redirect the user to one or more third-party trackers before redirecting the user to the intended destination. Figure~\ref{fig:uid_smuggling} describes this process. When a user clicks on a link on a website \emph{A}, the originator page itself or a tracker on the page--through a script--decorates the URL by adding the originator's user identifier (UID) as a query parameter. The user then passes through zero or more redirectors which are invisible to him. Each of these redirectors can get the UID from the query parameter and has permission to store it in a first-party cookie under the redirector's domain. Finally, the user is sent to the destination website B, and the redirector can forward or not to website \emph{B} the UID it received from \emph{A}. All the trackers on website B will be able to read the UID from the query parameter and know that it was the UID sent by the originator (through request headers). 

UID smuggling is more powerful than bounce tracking. Trackers using UID smuggling regain the ability to share UIDs across websites with different domains and can circumvent restrictions from partitioned cookie storage spaces~\cite{randall2022measuring}. For example, they can link the user's visits to the website \emph{A} with the user's visits to website \emph{B} and the user's activity that goes through its site (through redirects) since they can all be linked to the same user ID. 
In addition, UID smuggling can help other trackers on website \emph{B} (and website \emph{A}) to link users' browsing activity across all the websites that received the UID as a query parameter.

\section{Measurement methodology}
\label{sec:metho}
We develop a measurement methodology to capture network flows when clicking on an ad from a search engine results page. 
Using multiple crawlers, we simulate a large number of search engine queries in order to collect a sample of information flows per search engine. For each request, we collect the cookies created, the locally stored values, and the web request sent by the browser. 
In addition, we rely on several open-source datasets to detect web requests to online trackers. 
We consider five main search engines: Google\footnote{\url{https://www.google.com/}}, Bing\footnote{\url{https://www.bing.com/}}, \DDG\footnote{\url{https://duckduckgo.com/}}, \SP\footnote{\url{https://www.startpage.com}}, and Qwant\footnote{\url{https://www.qwant.com/}}. We use Google and Bing as baselines to compare with the other three, which claim to have higher privacy standards and protective measures in place.

\subsection{Crawling system}

Each crawling iteration begins at a search engine's main page, where our system will type a query and access the search engine results page. 
Next, it chooses one of the displayed ads to click on to access its destination website. Then, the navigation path passes through zero or more redirectors before landing on the ad's destination website. The redirectors are invisible to the user but can be identified through an analysis of network requests initiated by the browser. Each of these redirectors can read the query parameters added by the search engine or other intermediaries and store them locally or send them to other third parties. The system records all first-party and third-party cookies, local storage values, and web requests at each step. We run each iteration in a new browser instance to ensure no stale data is cached from previous iterations.

Depending on the search engine, ads are either part of the main page or are loaded through an iframe. 
We use scrapping techniques to detect them and rely on several HTML elements' attributes. 
For instance, all ads on \SP are inside an HTML element titled "Sponsored Links". 
Moreover, we use hyperlink values to detect Google ads since they all link to "\url{www.googleadservices.com/*}".

Our system prioritizes ads with landing domains it has not visited yet, aiming to maximize the number of different destination websites. Each time a crawler clicks on an ad, our system adds the domain of its landing URL to the list of visited websites. In the subsequent iterations, the crawler first extracts the landing domains of all the displayed ads. The landing domains are included within the HTML objects of the advertisements on all search engines. The crawler then gives preference to click on ads leading to domains that have not been encountered in the list of visited websites.

We reproduced these steps for 500 search queries on the five search engines. We randomly choose them from Google Trends~\cite{google_trends} and movie titles from MovieLens~\cite{movie_lens}. All iterations were performed in "accept" cookies mode. Table~\ref{tab:high_level_stats} represents the number of different search queries we typed,
the number of different destination pages we landed on, and the number of different domain paths we collected for each search engine.

\begin{table}
\caption{Number of search queries, destination websites, and redirection paths.}
\label{tab:high_level_stats}
\begin{tabular}{|l|l|l|l|l|}
\hline
& \# Queries & 
\begin{tabular}[c]{@{}l@{}} \# Different \\ destination \\websites\end{tabular}
& \begin{tabular}[c]{@{}l@{}}\# Different \\redirection\\ paths\end{tabular} 
\\ \hline           
Bing       & 500 & 98 & 131 \\ \hline
Google     & 500 & 102 & 134 \\ \hline
\DDG & 500 & 56 & 94 \\ \hline
\SP  & 500 & 60 & 107\\ \hline
Qwant & 500 & 60 & 88 \\ \hline
\end{tabular}
\vspace{-4mm}
\end{table}

We implemented our system using Puppeteer~\cite{puppeteer} to automate visiting search engines' websites, typing search queries, detecting and clicking on one of the displayed ads, and waiting for 15 seconds on the ad's destination website. We reproduce these steps multiple times from the same IP address for each search engine. To reduce the chance of being identified as bots, we use puppeteer-extra-plugin-stealth~\cite{puppeteer-extra-plugin-stealth}. This plugin applies various techniques to make the detection of headless Puppeteer crawlers by websites harder. 

Puppeteer allows us to record cookies and local storage for each request. However, it does not guarantee that it can attach request handlers to a web page before it sends any requests~\cite{randall2022measuring}. Hence, detecting and collecting web requests only using Puppeteer might cause losing some of them. We use a Chrome extension alongside Puppeteer crawlers to record web requests during all the crawling time. We do not observe a significant difference between web requests recorded by crawlers and web requests recorded by the extension. In median, the crawlers recorded 97\% of the requests recorded by the extension. The code of the crawling system and the dataset are available at~\url{https://github.com/CHOUAKIsalim/Search_Engines_Privacy}.

\subsection{Detection techniques}
\label{sec:methodology:techniques}

\vspace{2mm}
\noindent \textbf{Detection of trackers:} We use URL filtering to detect web requests to online trackers. 
We use filter rules from two open-source lists: EasyList~\cite{easylist} and EasyPrivacy~\cite{easyprivacy}. EasyList is the most popular list to detect and remove adverts from webpages and forms the basis of many combination and supplementary filter lists~\cite{easy_list_2}. EasyPrivacy is a supplementary filter list that detects and removes all forms of tracking from the internet, including tracking scripts and information collectors~\cite{easy_list_2}. These filter lists are used by extensions that aim to remove unwanted content from the internet, like AdBlock and uBlock. We combined and parsed these lists using adblock-rs~\cite{adblock_parser} and obtained \numprint{86488} filtering rules. 

In addition, we use the Disconnect Entity List~\cite{disconnect_entity_list} to get the entities of online tracker domains. It is a dictionary where keys represent entities such as Google, Microsoft, and Facebook, and values represent the web domains that belong to each entity. Hence, to get the entity of a tracker, we iterate over all values and search to what entity is the tracker domain associated with. This list contains \numprint{1449} entities and \numprint{3371} related web domains.

\vspace{2mm}

\noindent \textbf{Detection of bounce tracking:} 
We classify an instance as bounce tracking when an advertisement's destination link is altered to pass through one or more redirectors. To construct the redirection sequence, we trace the series of URLs the browser navigates through after clicking an ad and prior to reaching the advertisement's intended landing page. We further validate the redirection sequence by examining the HTTP response headers, precisely the 'Location' and 'status code' headers. These headers divulge the redirection process, as the 'Location' header contains the new redirection URL, and status codes such as '301 Moved Permanently,' '302 Found,' '307 Temporary Redirect,' and '308 Permanent Redirect' indicate the occurrence of redirection~\cite{redirection_codes}.

\vspace{2mm}
\noindent \textbf{Detection of  UID smuggling and user identifiers:}
To detect UID smuggling, we need to differentiate between query parameters that represent user identifiers and non-tracking query parameters such as session identifiers, dates, and timestamps.
We consider all query parameters, localStorage, and cookie values. We call them tokens. There are \numprint{6971} unique tokens in our dataset. We perform the following filtering, which is similar to the one performed by Randall et al.~\cite{randall2022measuring}:

\vspace{-1.5mm}

\begin{trivlist}
\item (i) Each iteration is executed in a new browser instance; hence, user identifiers should not be shared across browser instances. We discard tokens that are the same across all or a subset of browser instances. 
\item (ii) For each browser instance and search query, we analyze the tokens resulting from the URLs of all ads that appear on the results page (which are usually in the form of \url{googleadservices.com/..../aclk?..cid=CAESbeD2ZWCwqFv3e-2k_....}). We discard tokens with different values for the different ad URLs as they likely correspond to ad identifiers. 
\item (iii) To detect session identifiers, we store the profile of each iteration in a separate directory and execute an extra iteration per browser instance one day later to see which values of cookies/parameters change. We discard tokens with different values in the two iterations as they are more likely session identifiers. 
\item (iv) Similar to~\cite{randall2022measuring}, we use programmatic heuristics to discard particular values. We discard tokens that appear to be timestamps (values between June and December 2022 in seconds and milliseconds), tokens that appear to be URLs, tokens that constitute one or more English words (\cite{words_library}), and tokens that are seven characters long or less. 
\end{trivlist}

After using these filters, we are left with \numprint{1942} tokens. We manually investigated them and observed a non-negligible number of false positives. Hence, we manually filtered the remaining tokens and removed those composed of any combination of natural language words, coordinates, or acronyms. In the end, we are left with \numprint{1258} user-identifying tokens, which we consider to be user identifiers.

\section{Results}
\label{sec:results}

This section presents the results of applying the presented methodology to the five selected search engines. We measure how users' privacy is affected before, during, and after clicking on a search ad. We find that the advertising systems on all evaluated search engines result in privacy harm, even for search engines that market themselves as privacy-respecting. We find that how, and to what degree, user privacy is harmed varies across each evaluated system.

The rest of this section proceeds as follows. Section~\ref{sec:results:before} begins by presenting measurements of how user privacy is impacted \emph{before} users click on an ad (\IE{} after the user has received answers to their search query, but before the user
clicks on an advertisement contained among or alongside the search results).
Section~\ref{sec:results:during} presents measurements of how user privacy is effected \emph{during} clicking on an advertisement (\IE{} after the user has clicked on an advertisement, but before the user arrives at the advertisement's destination). Finally, Section~\ref{sec:results:after} gives measurements of how user privacy is affected \emph{after} clicking on an advertisement (\IE{} after the user has arrived at the final destination of the advertisement link, and scripts are executed on the advertiser's website).

\subsection{Before clicking on an ad}
\label{sec:results:before}
We first present measurements of how the advertising systems used by popular search engines affect user privacy before a person has clicked on any advertisement. At this point in the process, the user has submitted a query to the search engine and received a results page. The returned results include at least two types of links: ``organic results'' (\IE{} websites that contain content the search engine thinks relates to the query) and ``paid results'' (\IE{} advertisements that the search engine has been paid to show to users).

This subsection presents measurements of how user privacy is impacted 
before the user has clicked on a search advertisement. Since a user will only click on a fraction of the advertisements they are presented with, users will be effected by these ``before'' privacy harms more frequently than the privacy harms presented in later subsections.

\subsubsection{First-party reidentification}
\label{sec:results:before:cookies}
We first measure whether search engines track or reidentify users across queries and visits. We find that the non-privacy-focused search engines (\IE{} Bing and Google) track users across visits and are able to link different search queries to the same user who made those queries. The privacy-focused search engines, on the other hand, do not appear to attempt to reidentify users across visits or queries, aligning with the claims made in their privacy policies (see Section~\ref{sec:background:se}). We measured whether search engines are able to reidentify users across queries and visits by looking for whether search engines stored unique user identifiers in the browser's first-party storage (\EG{} cookies, localStorage). Specifically, we inspected the DOM storage area for each site and looked for stored values that appeared to be unique identifiers, using the heuristics described in Section~\ref{sec:methodology:techniques}. We observed that Google and Bing did store such user identifiers; the other search engines did not.

We note that some privacy-focused search engines \emph{did} store other values in first-party storage, but that they were used for purposes other than user identification (\EG{} client-side storage of user preferences).

\subsubsection{Requests to trackers}
\label{sec:results:before:trackers}
We also measured whether search engines harmed user privacy by communicating with trackers when presenting advertisements. We did not observe any search engine including resources from, or making network requests to, known trackers.

We checked for communication with known trackers by i. recording the URLs of all the network requests made by the browser when rendering the search results, and ii. checking those URLs against popular filter lists (as described in Section~\ref{sec:methodology:techniques}). These URLs comprise both the sub-resources (\EG{} scripts, images, videos) loaded by the results page and the third-party requests made using the Web networking APIs (\EG{} XMLHttpRequest, \texttt{fetch()}, web sockets).

We note that we were only able to measure the client-side network behavior of each search engine, and could only observe whether the search engine pages themselves were sharing information with known trackers. We were not able to measure how or if each search engine communicates with trackers on the server-side.

\subsection{When clicking on an ad}
\label{sec:results:during}
Next, we measure how user privacy is affected after the user clicks on an ad, but before the user has arrived at the ad's destination (usually, a page controlled by the party placing the advertisement). This step of the process involves systems run by both the search engine itself and the advertising platform paying for the ad.

During this stage, the advertising system may try and accomplish several goals, including fraud detection (\IE{} attempting to detect if the ``click'' was the result of an automated system, intending to increase how much the advertiser pays the search engine) and user profiling (\IE{} recording information about the user clicking the ad to combine with existing user profiles). Simultaneously, the search engine may use this step to try and achieve other goals, including quality of service measurements (\IE{} ensuring that advertisements render correctly) or additional user profiling (\IE{} recording which ad the user clicked to ``enrich'' whatever information the search engine may have about the user).

We find that the measured search engines vary widely in how they treat user privacy when the user clicks on an ad. However, we also find that the advertising systems engage in privacy-harming behaviors and share user identifying information with third parties across all measured search engines, despite the privacy-focused branding adopted by some search engines.

\subsubsection{Search engine page behaviors}
\label{sec:results:during:first}
First, we measured what behaviors the search engine's page engages in \emph{after} the user clicks on an ad but \emph{before} the browser begins navigating away from the search engine's page (and towards the advertisement's destination page). These behaviors might be things like recording which advertisement the user clicked on or how long the user waited before clicking, and are implemented with browser APIs like ``onclick'' handlers and ``ping'' attributes~\cite{ping_attribute}.

We measured each search engine's post-click behaviors by recording what network requests happened on the page after each advertisement was clicked on. We find that all search engines record additional information about the user and/or the user's click, after the user has clicked on an ad. 

\paragraph{Bing} 
Clicking on an advertisement on Bing results in additional first-party (\IE{} within Bing) network requests. In all iterations, clicking caused a request to be sent to \url{https://bing.com/fd/ls/GLinkPingPost.aspx}. These requests included several query parameters, including the clicked ads' destination websites. Furthermore, these requests include user identifiers, for instance, communicated in the \texttt{MUID} cookie --A cookie identifying unique web browsers visiting Microsoft sites-~\footnote{\url{https://learn.microsoft.com/en-us/clarity/cookie-list}}. 

\paragraph{Google}
Clicking on ads on Google results in additional first-party web requests. In all cases, the browser sends POST web requests to \url{https://google.com/gen_20?}. These requests include user identifier values communicated in cookies such as \texttt{NID} and \texttt{AEC}~\footnote{\url{https://policies.google.com/technologies/cookies}}. 

\paragraph{\DDG}
Clicking on an advertisement on \DDG{} results in additional first-party network connections to \url{https://improving.duckduckgo.com}. These requests include several query parameters, such as the search query, the ad provider (Bing in all cases), and the destination URL of the clicked ad. Next, the browser sends an additional network request that fetches a JavaScript file served from~\url{https://duckduckgo.com/y.js}. This request includes several query parameters containing information about the ad and the link to which the user should be redirected (link to Bing servers). We note that none of the query parameters nor the cookies sent with these web requests matched our user heuristics for user identifiers.

\paragraph{Qwant}
When clicking on an advertisement on Qwant, a first request is sent to \url{https://qwant.com/action/click_serp}, including information about the user's browser, such as the type of the device and the browser language, along with the search query. Furthermore, this request contains information on the clicked ad (\EG{} its position on the results page and the destination website). Then, another request is sent to \url{https://api.qwant.com/v3/redirect/}, including the URL to direct the user to. These two connections do not include user identifiers as query parameters nor as cookies values.

\paragraph{\SP}
Clicking on an advertisement on \SP{} results in an additional first-party request to \url{https://startpage.com/sp/cl}. This request includes information about the position of the clicked ad on the results page, but does not include the ad's destination URL. Similar to \DDG{} and Qwant, requests to StartPage servers do not include user identifiers. 

In summary, we find that all search engines, traditional and privacy-focused alike, record information about users' ad clicks. They all collect data about the clicked ad, such as its position on the results page or destination URL. However, only traditional search engines (Google and Bing) include user identifiers with web requests to their servers. 

\subsubsection{Navigation Tracking}
\label{sec:results:during:paths}
Next, we measure whether the advertising systems in search engines engage in navigation-based tracking, a technique for tracking users that circumvents browser privacy protections by directing a user through otherwise unrelated sites. Section~\ref{sec:background:nav} provides a high-level summary of how navigating tracking works and why it is an effective method of circumventing tracking protections in many browsers. We find that most of the search engines in our data set engage in navigation-based tracking at least some of the time. Further, we find that the \emph{privacy-focused search engines engage in navigation-based tracking for the majority of placed ads.}

We measure the navigation tracking we observed on the selected search engines in three dimensions: i. the distribution of how many sites the user is ``bounced'' through when they click on an ad on each search engine, ii. how many different organizations a user is exposed to during navigation tracking episodes (distinct from the number of pages or domains), and iii. the distribution of the number of sites in the redirection path that store user-identifying cookies. 

\begin{figure}[t!]
  \centering
  \includegraphics[width=\linewidth]{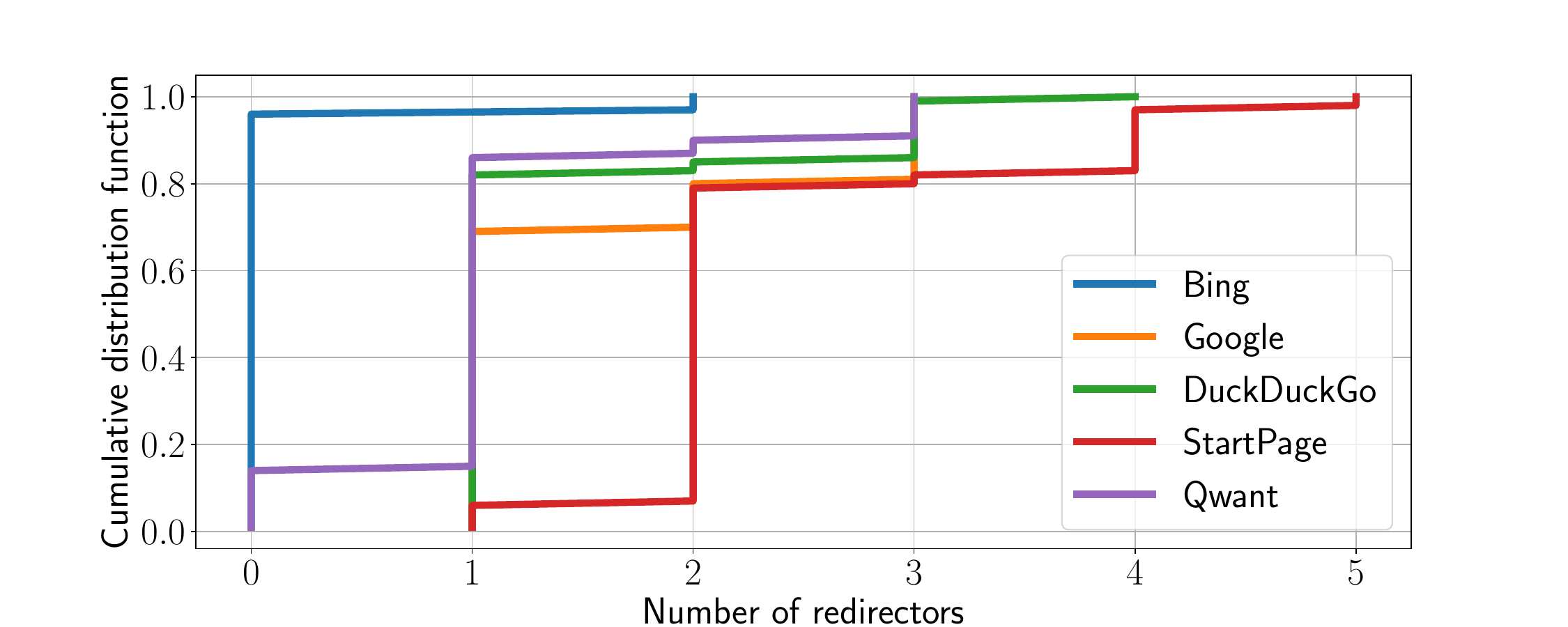}
  \caption{CDF of the number of different redirectors for Bing, \DDG, Google, and StartPage.}
  \label{fig:number_of_redirectors}
\vspace{-4mm}
\end{figure}

\paragraph{Number of sites visited} 

Figure~\ref{fig:number_of_redirectors} presents the distribution of the number of different sites (\IE{} $eTLD+1$) each search engine directs the user through when clicking on an ad.
We observe that clicking on an ad on Bing generally results in being redirected through the fewest number of sites (96\% of ad clicks on Bing result in no other site being visited except for Bing and the final destination site). 
Clicking on sites on \DDG, Google, and Qwant typically results in visiting one other site (respectively, 82\%, 69\%, and 72\% of clicks result in an intermediate navigation to a site different than the search engine and the ad's destination). 
Clicking on ads on \SP{} resulted in (on average) visiting the largest number of different sites (93\% of clicks resulted in visiting at least two sites other than \SP{} and the ad's destination).

\paragraph{Number of organizations visited}
However, we note that all redirections are not equal in their privacy impact; the marginal privacy harm is generally much lower if a site redirects the user between two sites the company owns, versus the user being redirected between two sites owned by unrelated companies. 
More concretely, there is little-to-no additional privacy harm if Google bounces a user---and passes information about the user---from \url{google.com} to \url{googleadservices.com}, while there \emph{is} privacy harm if Google bounces a user--and the user's information---from \url{google.com} to \url{facebook.com} (\IE{} Facebook learns new information they otherwise would not learn).

Understanding the privacy harm of navigation tracking requires considering \emph{which} sites the user is being ``bounced'' between. Table~\ref{tab:most_common_navigation_path} presents the five most common redirection paths for each search engine, and Table~\ref{tab:most_common_redirectors} in the appendix presents the most common sites in the redirection paths. Moreover, we group redirectors' domains by the organization to which they belong using the Disconnect Entity List~\cite{disconnect_entity_list}. Table~\ref{tab:redirectors_per_organization} presents the fraction of navigation paths that include a website from each organization across all search engines. 

We observe that the impact of navigation tracking differs widely between search engines. On one hand, the navigation tracking that occurs from clicking on ads on Google results in little additional privacy harm; the most commonly immediately visited sites are also operated by Google (\IE{} \url{googleadservices.com} and \url{ad.doubleclick.com}). On the other hand, we find that navigation tracking significantly harms user privacy on privacy-branded search engines. In all three cases, users are either usually directed to Bing sites (100\% and 76\% of the time for \DDG{} and Qwant, respectively) or Google sites (100\% of the time for \SP{}). 

While these results are alarming---since these are search engines advertising that they are privacy-preserving ---they are not inexplicable. \DDG{} and Qwant rely on Bing to provide search ads, and \SP{} relies on Google.

\begin{table*}
\caption{Top five most common navigation domain paths when clicking an ad for each search engine.}
\label{tab:most_common_navigation_path}

\begin{tabularx}{\textwidth}{|l|X|l|}
\hline
\textbf{Search engine} & \textbf{Domain paths} & \textbf{Frequency} \\ \hline \hline
\multirow{4}{*}{Bing} & bing.com - destination & 96\% \\
 & bing.com - clickserve.dartsearch.net - ad.doubleclick.net - destination & 3\% \\ 
 & bing.com - t23.intelliad.de - 1045.netrk.net - destination	 & 1\% \\ \hline
\multirow{5}{*}{Google} & google.com - googleadservices.com - destination & 69\% \\ 
 & google.com - googleadservices.com - clickserve.dartsearch.net - ad.doubleclick.net - destination & 17\% \\ 
 & google.com - googleadservices.com - pixel.everesttech.net - ad.doubleclick.net - destination & 4\% \\ 
 & google.com - googleadservices.com - monitor.clickcease.com - destination	& 4\% \\
 & google.com - googleadservices.com - monitor.ppcprotect.com - destination & 2\% \\ \hline
\multirow{5}{*}{\DDG} & duckduckgo.com - bing.com - destination & 82\% \\ 
 & duckduckgo.com - bing.com - clickserve.dartsearch.net - ad.doubleclick.net - destination & 14\% \\
& duckduckgo.com - bing.com - 6102.xg4ken.com - destination & 2\% \\ 
& duckduckgo.com - bing.com - clickserve.dartsearch.net - ad.doubleclick.net - tpt.mediaplex.com - destination & 1\% \\
& duckduckgo.com - bing.com - pixel.everesttech.net - destination	 & 1\% \\ \hline
\multirow{5}{*}{StartPage} & startpage.com - google.com - googleadservices.com - destination &	73\% \\
& startpage.com - google.com - googleadservices.com - clickserve.dartsearch.net - ad.doubleclick.net - destination	 & 17\% \\
& startpage.com - google.com - destination & 6\% \\
& startpage.com - google.com - googleadservices.com - 6008.xg4ken.com - destination	& 1\% \\
& startpage.com - google.com - googleadservices.com - clickserve.dartsearch.net - ad.doubleclick.net - monitor.ppcprotect.com - destination &	1\% \\ \hline
\multirow{5}{*}{Qwant}
& qwant.com - bing.com - destination	& 66\% \\
& qwant.com - destination	& 14\% \\
& qwant.com - bing.com - clickserve.dartsearch.net - ad.doubleclick.net - destination & 10\% \\
& qwant.com - track.effiliation.com - destination	& 3\% \\
& qwant.com - click.linksynergy.com - destination	& 3\% \\
\hline

\end{tabularx}
\vspace{-2mm}
\end{table*}

{\footnotesize
\begin{table}
\caption{Fraction of navigation paths that include a website from each organization across all search
engines.}
\label{tab:redirectors_per_organization}
\begin{tabular}{|l|l|l|l|l|l|}
\hline
\textbf{} & \textbf{Bing} & \textbf{Google} & \textbf{\DDG} & \textbf{\SP} & \textbf{Qwant} \\ \hline \hline
\textbf{Adobe} &    0\% &           \textbf{4\%} & \textbf{1\%} &  0\% &          \textbf{1\%} \\ \hline
\makecell[l]{\textbf{Conversant}\\\textbf{Media}} & 0\% &    0\% &          \textbf{1\%} &  0\% &          0\% \\ \hline
\makecell[l]{\textbf{DuckDuck}\\\textbf{Go}} & 0\% &         0\% &          100\% &         0\% &          0\% \\ \hline
\textbf{Facebook} & 0\% &           0\% &          0\% &           \textbf{1\%} & 0\% \\ \hline
\textbf{Google} & \textbf{3\%} &    100\% &       \textbf{15\%} &  100\% &        \textbf{11\%} \\ \hline
\textbf{Kenshoo} & 0\% &            \textbf{2\%} & \textbf{2\%} &  \textbf{1\%} & 0\% \\ \hline
\textbf{Microsoft} & 100\% &        0\% &          100\% &         0\% &          79\% \\ \hline
\textbf{Nielsen} & 0\% &            0\% &          0\% &           \textbf{1\%} & 0\% \\ \hline
\textbf{PPCProtect} & 0\% &         \textbf{2\%} & 0\% &           \textbf{1\%} & \textbf{1\%} \\ \hline
\textbf{Qwant} & 0\% &              0\% &          0\% &           0\% &          100\% \\ \hline
\textbf{Rakuten} & 0\% &            0\% &          0\% &           0\% &          \textbf{3\%} \\ \hline
\textbf{StartPage} & 0\% &          0\% &          0\% &           100\% &        0\% \\ \hline
\textbf{Unknown} & \textbf{4\%} &  \textbf{23\%} & \textbf{15\%} & \textbf{19\%} & \textbf{16\%} \\ \hline

\end{tabular}

\end{table}
}
\begin{figure}[t!]
  \centering
  \includegraphics[width=\linewidth]{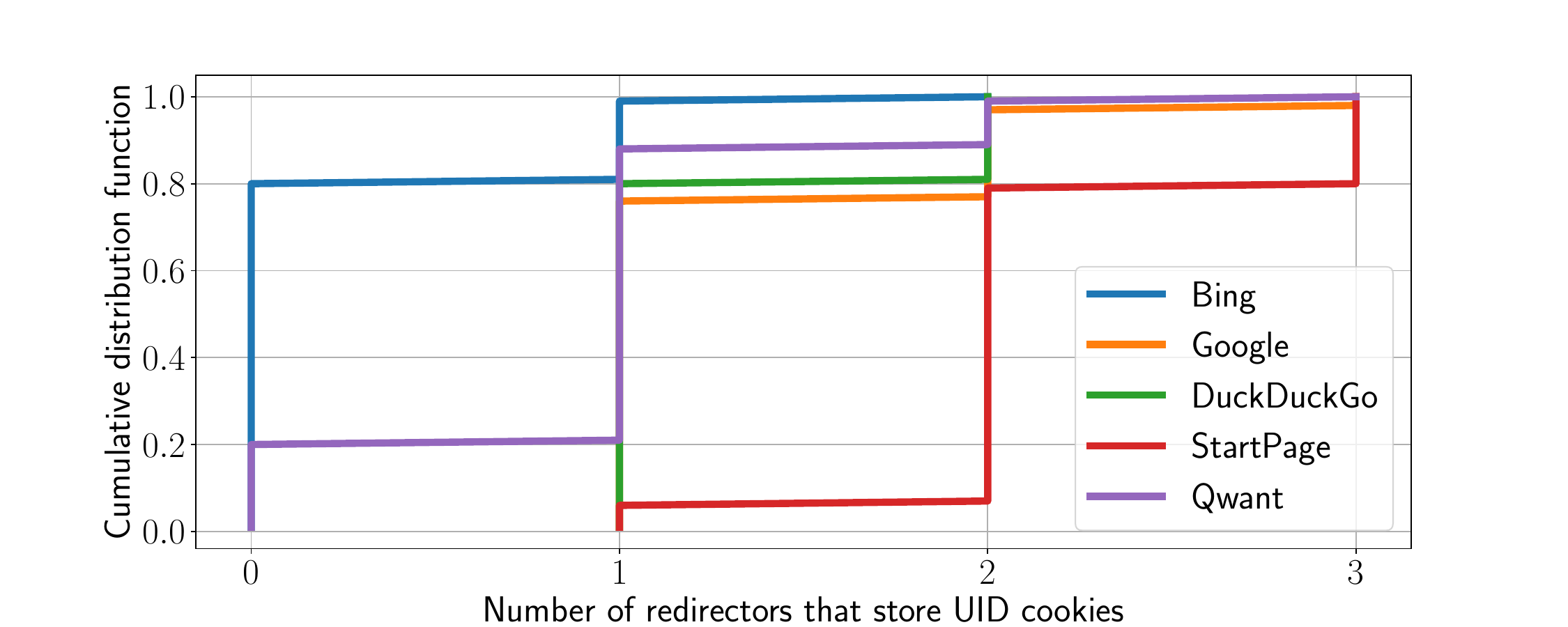}
  \caption{CDF of the number of different redirectors that store UID cookies for Bing, \DDG, Google, and StartPage.}
  \label{fig:number_of_redirectors_uid_cookies}
  \vspace{-4mm}
\end{figure}

\paragraph{Number of sites that identify users}

The extent of privacy harm resulting from bounce tracking depends on two key factors: the behavior of the redirector (\IE{} whether the redirector stores user-identifying cookies) and the type of cookie storage used by the browser (flat or partitioned). The lowest level of privacy harm occurs when the redirector does not store any user-identifying cookies. In this case, the redirector can infer the source and destination of the navigation event (\IE{} the search engine and the ad's website). However, if the user navigates through the same redirector multiple times, the redirector cannot aggregate the tracking data from different visits to the same user.

In contrast, if the redirector sets UID cookies on users' browsers, it can combine tracking data each time the user bounces through it. Specifically, if a user clicks on multiple ads on the same search engine and is redirected through the same redirector each time, the redirector can aggregate all the websites the user has visited. Moreover, if the user's browser has flat cookie storage, the redirector can potentially aggregate the user's activity and match it to the same user instance on every website where the redirector has a script.

\begin{table*}[t!]
\caption{Redirectors that store UID cookies.}
\label{tab:redirectors_creating_cookies}
\begin{tabular}{| l | l | l | l | l |}
\hline
\textbf{Bing} & 
\textbf{Google} & 
\textbf{\DDG} & 
\textbf{\SP} & 
\textbf{Qwant} 
\\\hline \hline
\makecell[l]{ad.doubleclick.net \\(3\%)} &
\makecell[l]{googleadservices.com \\(98\%)} &
\makecell[l]{bing.com \\(95\%)} & 
\makecell[l]{google.com \\(100\%)} & 
\makecell[l]{bing.com \\(78\%)} 
\\ \hline
\makecell[l]{t23.intelliad.de \\(1\%)} & 
\makecell[l]{ad.doubleclick.net\\(21\%)} &
\makecell[l]{ad.doubleclick.net\\(14\%)} &
\makecell[l]{googleadservices.com \\(94\%)} &
\makecell[l]{ad.doubleclick.net \\(11\%)} 
\\ \hline
\makecell[l]{1045.netrk.net \\(1\%)} & 
\makecell[l]{pixel.everesttech.net \\(4\%)} &  
\makecell[l]{6102.xg4ken.com \\(2\%)} &
\makecell[l]{ad.doubleclick.net \\(18\%)} & 
\makecell[l]{click.linksynergy.com \\(3\%)} 
\\ \hline
 & 
 \makecell[l]{monitor.ppcprotect.com \\(2\%)} &
 \makecell[l]{pixel.everesttech.net \\(1\%)} & 
 \makecell[l]{6008.xg4ken.com \\(1\%)} & 
 \makecell[l]{pixel.everesttech.net \\(1\%)} 
 \\ \hline
 & 
 \makecell[l]{3825.xg4ken.com \\(2\%)} 
 & 
 & 
 & 
 \makecell[l]{monitor.ppcprotect.com	\\(1\%)} 
 \\ \hline
 &
 &
 &
 &
 \makecell[l]{tracking.deepsearch.adlucent.com \\(1\%)} 
 \\ \hline

\end{tabular}
\vspace{-2mm}
\end{table*}

Figure~\ref{fig:number_of_redirectors_uid_cookies} presents the distribution of the number of different redirectors in the navigation paths that store UID cookies for each search engine, and the Table~\ref{tab:redirectors_creating_cookies} lists these redirectors that store UID cookies on users' browsers. Our observations indicate that the level of privacy harm resulting from bounce tracking varies considerably across different search engines. While the navigation tracking that occurs when users click on ads on traditional search engines appears to cause little privacy harm, as users are identified by sites operated by third-party entities in only 4\% and 8\% of navigation paths for Bing and Google, respectively. In contrast, for the three privacy-branded search engines, users are identified by sites operated by third-party entities in most cases. Precisely, more than 95\% of users clicking on ads on \DDG{}, \SP{}, or Qwant are identified by Bing, and Google identifies users clicking on ads on StartPage in 100\% of cases. As a result, Google and Bing might associate the destination website visited by the user through the advertisement to the user profile, especially if the user's browser has flat-cookie storage.

\subsection{After clicking on an ad}
\label{sec:results:after}
Finally, we measure how user privacy is impacted once the user has ``finished'' clicking on a search ad and has arrived at the advertiser's page. We measure how the search engine/advertiser relationship effects user privacy in two ways: first, by measuring whether advertisers include trackers or other known-privacy-harming resources, and two, by measuring if and what kinds of information the search engine's advertising system provides to the advertiser (in the form of user-describing query params). This first measure relates to whether the search engine requires advertisers to abide by privacy-respecting practices; the latter measure relates to whether search engines' advertising systems collude with advertisers to aid advertisers in profiling visitors.

Redirectors in navigation paths can aggregate more data about the user's behavior if they have scripts on the ads' destination websites. 
For this, they need to match users using either third-party cookies if they are enabled by the browser or UID smuggling. 
We investigate whether redirectors can aggregate users' activity on ads destination websites by analyzing online trackers, whether they receive UID as query parameters, and whether they store them. 
We recorded these requests by keeping the crawlers on the ads' destination pages for 15 seconds for all iterations. 

\subsubsection{Requests to online trackers}
We first measure whether search engines protect their users by
requiring advertisers to be privacy-protecting. We measure this by loading the website each clicked search advertisement leads to, recording the URLs of all sub-resources and network requests made when loading and executing the page, and comparing those URLs against \EL{} and \EP{}.

We find that 93\% of the web pages users are taken to when they click on ads on both ``standard'' and ``privacy-focused'' contain many privacy-harming resources. 
Broken down by search engine, we observed 277, 218, 326, 437, and 260 different tracker third parties over all iterations, and a median of 9, 11, 6, 8, and 6 different online trackers per iteration for Bing, Google, \DDG, \SP, and Qwant, respectively.

In order to understand which companies track users on ad destination pages, we group the domains that observed tracking resources are served from by ``entity'' using the Disconnect Entity List~\cite{disconnect_entity_list}
For example, using the entity list, we group tracker resources served from the domains \url{google.com} and \url{doubleclick.com} to the same entity (\IE{} Google). Table~\ref{tab:top_domain_entites_after_reaching_destination} presents the top entities of trackers we observed on ad destination pages.
For instance, we see that Google is the top entity for online trackers on destination pages for \SP (36\%), and we saw that all \SP redirection paths go through Google servers. 
Hence, if the browser implements a flat cookies storage, Google can match the \SP user on the ads destination website and aggregate data about his activity on it in 36\% of the cases. We make the same observation for Microsoft trackers on Qwant (4.3\%).

\begin{table*}[t!]
\caption{Top entities of online trackers reached by crawlers on each search engine.}
\label{tab:top_domain_entites_after_reaching_destination}
\begin{tabularx}{\textwidth}{| X | X | X | X | X | X |}
\hline
\textbf{Bing} & 
\textbf{Google} & 
\textbf{\DDG} & 
\textbf{\SP} & 
\textbf{Qwant} 
\\\hline \hline

unknown (32.0\%) & unknown (34.8\%) & unknown (29.5\%) & Google (36.0\%) & Google (26.3\%) \\ \hline

Google (24.4\%) & Google (28.7\%) & Google (21.8\%) & unknown (28.1\%) & Amazon (23.4\%) \\ \hline

Microsoft (13.8\%) & Microsoft (10.5\%) & Amazon (16.3\%) & Microsoft (4.3\%) & unknown (22.4\%) \\ \hline

Facebook (3.8\%) & Amazon (3.1\%) & Facebook (3.4\%) & Facebook (3.2\%) & Microsoft (4.2\%) \\ \hline

Criteo (2.4\%) & Criteo (2.5\%)	& Criteo (2.2\%) & Criteo (3.0\%) & Criteo (3.8\%) \\ \hline

\end{tabularx}
\vspace{-2mm}
\end{table*}

\subsubsection{User identifiers}
\begin{table}[t!]
\caption{Fraction of iteration where the ad's destination page received MSCLKID, GCLID and other UID attributes as query parameters.}
\label{tab:query_parameters_destination_page}
\begin{tabular}{|l|l|l|l|}

\hline
                           & MSCLKID                    & GCLID                     & \makecell[l]{other UID \\parameters}         \\ \hline
\multicolumn{1}{|l|}{Bing} & \multicolumn{1}{l|}{79\%} & \multicolumn{1}{l|}{12\%} & \multicolumn{1}{l|}{3\%} \\ \hline
Google                     & 0\%                       & 92\%                      & 8\%                      \\ \hline
\DDG                       & 66\%                      & 12\%                      & 6\%                      \\ \hline
\SP                  & 0\%                       & 92\%                      & 12\%                     \\ \hline
Qwant                      & 51\%                      & 8\%                       & 7\%          \\ \hline            

\end{tabular}
\vspace{-4mm}
\end{table}

Finally, we measure if the advertising systems of the search engines aid advertisers in tracking users across sites by transmitting unique identifiers (or other personal or otherwise individual values) across site boundaries through query parameters.

As discussed in Section~\ref{sec:methodology:techniques}, this technique is sometimes called UID smuggling and is a common technique trackers and sites use to circumvent browser privacy protections (such as blocking third-party cookies or partitioning browser storage). For example, if an advertiser places an ad for \url{https://site.example}, the advertising system might collude with the advertiser to allow the advertiser to profile the user by appending unique identifiers to the destination URL. The search engine's advertising system might, for example, append information the advertising system knows about the user to the advertiser's destination URL (creating a URL like \url{https://site.example?user_id=<id>}, so that the advertiser can learn more about the user, harming the user's privacy.

We measure whether search engines' advertising systems collude with advertisers to track users across sites by examining the query parameters the search engine (or other intermediate party in a navigation chain)
includes in the URL of the advertiser's destination page. We collect all of the query parameters in the destination ad URLs and extracted values that appeared to be unique identifiers using the heuristics described in Section~\ref{sec:methodology:techniques}.

We find that advertising systems collude with advertisers most of the time across all search engines, \emph{even private ones}. Clicking ads on all five search engines resulted in
user identifiers being passed to advertisers. We found user identifiers in query parameters in 80\%, 94\%, 68\%, 92\%, and 53\% for Bing, Google, \DDG, \SP, and Qwant, respectively. Most of these parameters are MSCLKID (Microsoft Click Identifier) or GCLID (Google Click Identifier), two \emph{unique identifiers} used for ad-click tracking. MSCLKID is added 
by Microsoft Advertising and GCLID 
by Google Ads when users click on their respective ads. Advertisers use these 
IDs to identify and track ad clicks; advertisers might store click-tracking first-party cookies to track actions taken after the ad click~\cite{msclkid, gclid, gclid_questions}. Table~\ref{tab:query_parameters_destination_page} represents the fraction of iteration where the web request to the ad's destination page included MSCLKID, GCLID, or other parameters. We can see that in search engines that use Microsoft advertising (\DDG and Bing), we find both MSCLKID and GCLID. However, in ones that use Google advertising (Google, \SP, and Qwant), we do not find MSCLKID.

Moreover, we investigate whether advertisers persist the UID query parameters they receive. We cross-reference values obtained from destination pages' first-party storage (e.g., cookies and localStorage) with the query parameters these pages receive. We find that MSCLKID values are persisted in 15\%, 17\%, and 1\% of cases for Bing, \DDG, and Qwant, respectively. As for GCLID, we find that a cookie is created in 5\%, 10\%, and 13\% of cases for Bing, Google, and \SP.

\section{Limitations}

Our measurement methodology has some limitations. First, we only look for user identifiers transferred in query parameters and do not detect them when they are transferred in other methods. For instance, previous work~\cite{randall2022measuring, webkit_trakcking_prevention_policy} found that trackers sometimes decorate their own URL in the document.referrer header with user identifiers and reads them on the destination page. Second, we run all our crawling iterations from the same IP address. Consequently, if some query parameters are IP address based, they will have the same value across all iterations, and thus we would not consider them as user identifiers. Finally, our results
are subject to variation based on the ads we selected and the search queries we used. Different search queries could potentially trigger distinct ads and lead to diverse advertisers, potentially exhibiting different behaviors. Nonetheless, our primary objective is to demonstrate the potential for third-party tracking when interacting with ads on private search engines.

\section{Related work}

Search engines and online tracking received a lot of research attention. We review studies closest to our work.

\textbf{Search engines.}
A first line of work has measured to which extent we can observe personalization in search engine results~\cite{arxiv.2211.11518,10.1145/2488388.2488435}  and ads~\cite{10.1145/1879141.1879152}.
For instance, Hannak et al.~\cite{10.1145/2488388.2488435} have developed a methodology for measuring personalization in search results, applied it to Bing, Google, and \DDG, and found that Bing results are more personalized than Google ones while they did not notice any personalization for DuckDuckGo. A second line of work has focused on solutions to protect users' privacy from search engines and prevent web profiling.
Castellà-Roca et al.~\cite{CASTELLAROCA20091541} presented a computationally efficient protocol that provides a distorted user profile to the search engine to preserve users' privacy. Finally, several studies have proposed privacy-preserving search-personalizing solutions for search engines. For instance, Shen et al.~\cite{10.1145/1273221.1273222} analyze various software architectures for personalized search and envision possible strategies with a client-sided personalization. Xu et al.~\cite{10.1145/1242572.1242652} suggest helping users choose the content and degree of detail of the profile information built by search engines. To the best of our knowledge, there is no study investigating the privacy properties of the advertising systems used on private search engines.

\vspace{2mm}
\textbf{Online tracking.}
Several works analyzed the usage of cross-site tracking techniques in the wild~\cite{Demir_2022}. Chen et al.~\cite{10.1145/3442381.3449837} propose a data flow tracking system to measure user tracking performed through first-party cookies. They found that more than 97\% of the websites they have crawled have first-party cookies set by third-party javascript and that on 57\% of them, there is at least one cookie containing a unique user identifier diffused to multiple third parties. Roesner et al.~\cite{10.5555/2228298.2228315} measured how user tracking occurs in the wild. They found that multiple parties track most commercial pages and estimate that several trackers can each capture more than 20\% of a user's browsing behavior. 
Koop et al.~\cite{Koop_indepth_evaluation} analyzed a dataset of redirection chains in the wild and found that 11\% of websites redirect to the same 100 top redirectors. Moreover, they demonstrate that these top redirectors could identify users on the most visited websites. Randall et al.~\cite{randall2022measuring} measure the frequency of UID smuggling in the wild and find that it is performed on more than 8\% of all navigations in their dataset. We use a similar method to identify user identifiers among all cookie values and query parameters by implementing automatic filtering followed by a manual inspection. All these studies were conducted in the wild, and to the best of our knowledge, no study focuses on navigational tracking techniques performed on search engines.

\section{Conclusion}

In this paper, we presented the first systematic study of the privacy properties of the advertising systems of five popular search engines: Two traditional ones, Google and Bing, and three private ones, \DDG, \SP, and Qwant. 
We investigated whether, and to which extent, search engines through their advertising systems, engage in privacy-harming behaviors that allow cross-site tracking.

Despite the privacy 
intentions and
promises of private search engines, 
our findings reveal the failure of privacy-focused search engines to fully protect users' privacy during ad interactions. Users on all measured search engines, including the privacy-focused ones, 
are subject to navigation-based tracking by third parties. 
We find that all search engines engage in bounce tracking when clicking on ads, where users are sent through several redirectors before reaching the ads' destination websites. 
While private search engines themselves do not engage in user tracking, their reliance on traditional advertising systems (Microsoft or Google) renders users susceptible to tracking by those systems. 
\emph{Although we cannot directly attribute this tracking to the search engines themselves, it is evident that they are enabling it through their reliance on Microsoft and Google's advertising systems.}

Inspecting the privacy policies of the search engines in light of our findings reveals interesting disparities. 
While our results demonstrate that Microsoft is capable of tracking \DDG users when they click on ads, \DDG asserts that Microsoft does not associate ad-click data with user profiles. 
On the other hand, Qwant, which also relies on Microsoft advertising for a significant fraction of its ads, do not document the utilization of ad-click data by Microsoft and whether it is used to enhance user profiles. 
Similarly, \SP explicitly states that clicking on ads subjects users to the data collection policies of other websites.

Our study highlights the need for increased attention to privacy protection within the advertising systems of search engines. 
One potential solution to protect users' privacy for private search engines would be to reduce their reliance on third-party advertising systems. Developing their own advertising platform could provide greater control over privacy practices, although the feasibility and complexity of such an approach remain uncertain. 
Alternatively, private search engines could collaborate with advertising systems such as Microsoft and Google, forging partnerships that proactively tackle privacy concerns. For instance, private search engines could negotiate agreements with the ad provider that prevent redirecting users who click on ads placed within private search engines to additional third parties. This approach would minimize the extent of third-party tracking, limiting it to the ad provider only. 
Moreover, search engines like \SP and Qwant could follow the lead of \DDG by seeking agreements with advertising systems to prevent the use of ad-click identifiers for user profile enrichment. These proactive steps would enhance user privacy while maintaining advertising partnerships with larger platforms.

\section*{ACKNOWLEDGMENTS}This research was supported in part by the French National Research
Agency (ANR) through the ANR-17-CE23-0014, ANR-21-CE23-0031-02, and MIAI@Grenoble Alpes ANR-19-P3IA-0003 grants and by the EU through the 
101041223, 101021377, and 952215 grants.

\bibliographystyle{ACM-Reference-Format}
\bibliography{links, references}

\footnotesize{
\begin{table*}[bp]
\caption{Most common redirectors (and their fractions) in domain navigation paths when clicking an ad on search engines. }
\label{tab:most_common_redirectors}

\begin{tabular}{|l|l|l|l|l|l|}
\hline
\makecell[l]{\textbf{Bing}} & 
\makecell[l]{\textbf{Google}} &
\makecell[l]{\textbf{DuckDuckGo}} &
\makecell[l]{\textbf{StartPage}} &
\makecell[l]{\textbf{Qwant}}  \\ \hline \hline
      
\makecell[l]{clickserve.dartsearch.net \\ (38\%)} &  
\makecell[l]{googleadservices.com \\ (65\%)} & 
\makecell[l]{bing.com \\ (74\%)} & 
\makecell[l]{google.com \\ (42\%)} &
\makecell[l]{bing.com \\(71\%)}  \\ \hline

\makecell[l]{ad.doubleclick.net \\ (37\%)} & 
\makecell[l]{ad.doubleclick.net \\ (14\%)} & 
\makecell[l]{clickserve.dartsearch.net \\ (11\%)} & 
\makecell[l]{googleadservices.com \\ (39\%)} & 
\makecell[l]{ad.doubleclick.net \\ (10\%)} \\ \hline

\makecell[l]{t23.intelliad.de \\ (13\%)} & 
\makecell[l]{clickserve.dartsearch.net \\ (13\%)} & 
\makecell[l]{ad.doubleclick.net \\ (11\%)} & 
\makecell[l]{clickserve.dartsearch.net \\ (7\%)} & 
\makecell[l]{clickserve.dartsearch.net \\ (9\%)}

\\ \hline

\makecell[l]{1045.netrk.net \\ (12\%)} & 
\makecell[l]{pixel.everesttech.net \\ (3\%)} &	
\makecell[l]{6102.xg4ken.com \\ (2\%)} & 
\makecell[l]{ad.doubleclick.net \\ (7\%)} &	
\makecell[l]{track.effiliation.com \\ (3\%)}

\\ \hline

& 
\makecell[l]{monitor.clickcease.com \\ (3\%)} & 
\makecell[l]{tpt.mediaplex.com \\ (1\%)} & 
\makecell[l]{6008.xg4ken.com \\ (1\%)} &
\makecell[l]{click.linksynergy.com \\ (2\%)} 
\\ \hline

& 
\makecell[l]{monitor.ppcprotect.com \\ (1\%)} & 
\makecell[l]{pixel.everesttech.net \\ (1\%)} & 
\makecell[l]{monitor.ppcprotect.com \\ (1\%)} & 
\makecell[l]{pixel.everesttech.net \\ (1\%)} 
\\ \hline

 & 
 \makecell[l]{3825.xg4ken.com \\ (1\%)} & & 
 \makecell[l]{t.myvisualiq.net \\ (1\%)} & 
 \makecell[l]{awin1.com \\ (1\%)}

 \\ \hline

 & & & 
 \makecell[l]{monitor.clickcease.com \\ (1\%)} & 
 \makecell[l]{zenaps.com \\ (1\%)}  \\ 
 
 \hline

& & & 
\makecell[l]{ad.atdmt.com \\ (1\%)} & 
\makecell[l]{deepsearch.adlucent.com \\ (1\%)}  \\ \hline
& & & & 
\makecell[l]{monitor.ppcprotect.com \\ (1\%)}  \\ \hline

\end{tabular}
\end{table*}
}

\section{Appendix}

\subsection*{Ethics}
Our experiments were conducted in a completely automated manner, without any human involvement or use of user data. Furthermore, the measurements we performed imposed minimal overhead on the well-resourced ad networks

\end{document}